\documentclass[conference,10pt]{IEEEtran}
\IEEEoverridecommandlockouts

\usepackage{cite}
\usepackage{booktabs}
\usepackage{amsmath,amssymb,amsfonts}
\usepackage{graphicx}
\usepackage{textcomp}
\usepackage{xcolor}
\usepackage{algpseudocode}
\usepackage{siunitx}
\DeclareSIUnit{\nothing}{\relax}
\DeclareSIUnit{\MAC}{MAC}
\usepackage{multirow}

\usepackage{algorithm}
\usepackage{url}

\usepackage{tikz}
\usetikzlibrary{shapes.misc}

\definecolor{Blue}{HTML}{0b0bfe}
\definecolor{Green}{HTML}{0c860c}

\newcommand{\numberincircleblue}[1]{%
  \tikz[baseline=(char.base)]{
    \node[shape=circle, draw=white, inner sep=1pt, fill=Blue] (char)
      {\color{white}\sffamily\bfseries\footnotesize #1};
  }%
}
\newcommand{\numberincirclegreen}[1]{%
  \tikz[baseline=(char.base)]{
    \node[shape=circle, draw=white, inner sep=1pt, fill=Green] (char)
      {\color{white}\sffamily\bfseries\footnotesize #1};
  }%
}

\def\BibTeX{{\rm B\kern-.05em{\sc i\kern-.025em b}\kern-.08em
    T\kern-.1667em\lower.7ex\hbox{E}\kern-.125emX}}

\title{Adaptive AI: Energy Efficient Multi-exit TinyML on Intelligent Vision Systems at the Edge}

\definecolor{somegray}{rgb}{0.5, 0.5, 0.5}
\newcommand{\darkgrayed}[1]{\textcolor{somegray}{#1}}
\makeatletter
\newcommand*\titleheader[1]{\gdef\@titleheader{#1}}
\AtBeginDocument{%
  \let\st@red@title\@title
  \def\@title{%
    \vskip-1.7em
    \bgroup\normalfont\large\centering\@titleheader\par\egroup
    \vskip0.4em\st@red@title}
}

\makeatother
\titleheader{\darkgrayed{This paper has been accepted for publication in the IEEE International Conference on Omni-Layer Intelligent Systems (COINS) \copyright 2026 IEEE.}}

\begin{document}

\author{
Luca~Crupi, Lorenzo~Lamberti, Alessandro~Giusti, and~Daniele~Palossi%
\thanks{L. Crupi, L. Lamberti, A. Giusti, and D. Palossi are with IDSIA, SUPSI, Lugano, Switzerland. Corresponding author: {\tt\small luca.crupi@supsi.ch}.}%
\thanks{L. Lamberti and D. Palossi are also with the IIS, ETH Z\"urich, Switzerland.}%

\thanks{This work has been partially supported by the Italy-Swiss EU Interreg project ALP-AI (grant number 0300290).}%
}


\maketitle
\IEEEpubid{\makebox[\columnwidth]{979-8-3195-0489-0/26/\$31.00~\copyright2026 IEEE\hfill}%
  \hspace{\columnsep}\makebox[\columnwidth]{ }}

\begin{abstract}
Traditional TinyML systems for edge devices achieve high accuracy by relying on fixed-depth models that require a constant number of multiply-accumulate (MAC) operations regardless of the input complexity. 
This approach wastes critical resources in battery-powered Internet-of-Things (IoT) devices and limits the real-time performance of edge cyber-physical systems.
Multi-exit execution schemes mitigate these issues and are widely used on high-end devices such as GPUs, but are rarely exploited on edge IoT devices because they require substantial rethinking given their strict memory and computational constraints.
We address these aspects by designing and deploying, on an ultra-low-power GWT GAP9 System-on-Chip (SoC), a novel multi-exit computational scheme, demonstrating it on a MobileNetV2 convolutional neural network (CNN) for the ImageNet-100 classification task.
Our approach introduces multiple exits at different CNN depths, each with a confidence-based gating mechanism that dynamically and autonomously decides whether to continue or stop inference.
Comparing our multi-exit strategy to the standard MobileNetV2 on a GAP9 SoC, we show a 41\% reduction in the average computational cost (from \textbf{\SI{313}{\mega\MAC}} to \SI{185}{\mega\MAC}), a 29\% lower inference time (from 49 to \SI{35}{\milli\second}), and an energy saving of 24\% (from 2.1 to \SI{1.6}{\milli\joule} per frame).
All these improvements come with a $\sim$1\% loss in accuracy compared to the full-depth MobileNetV2, which achieves 80.5\%.
Finally, comparing our adaptable multi-exit scheme with a third-party state-of-the-art adaptive CNN, also deployed on the GAP9, we achieve more than 2$\times$ its computational efficiency, increasing it from 8.1 to \SI{17.2}{MAC/cycle}.
\end{abstract}

\renewcommand{\IEEEkeywordsname}{Keywords}

\section{Introduction}
\IEEEpubidadjcol

Edge execution of efficient tiny machine learning (TinyML) workloads is fueling the advancement of battery-powered Internet-of-Things (IoT) sensors, wearable devices, and miniaturized robots~\cite{TinyML_survey_songhan, survey_edge, palossi_vo, wearable_tinyml, niculescuAICAS}. 
The forefront of this computational paradigm is represented by extreme edge computing~\cite{TinyML_survey_songhan}, where ultra-low-power (ULP), resource-constrained systems, based on sub-\SI{100}{\milli\watt} microcontroller units (MCUs), execute complex TinyML models, e.g., up to a few hundred million multiply-accumulate (MAC) operations~\cite{TinyML_survey_songhan, niculescuAICAS, crupi_pest_2024, bompani_bytetrack}, often under strict real-time constraints~\cite{crupi2024high, lamberti2024distilling, LambertiCereda}.
Embodied artificial intelligence (AI) systems and agentic AI at the edge exacerbate this scenario, requiring high-throughput, low-latency output to deliver critical information for controlling subsystems and actuators.
A notable example is given by \SI{}{\centi\meter}-scale autonomous robots running TinyML visual perception pipelines that feed high-frequency controllers~\cite{crupi2023sim, crupi2024high, LambertiCereda}.

\begin{figure}[t]
\centering
\includegraphics[width=1.0\columnwidth]{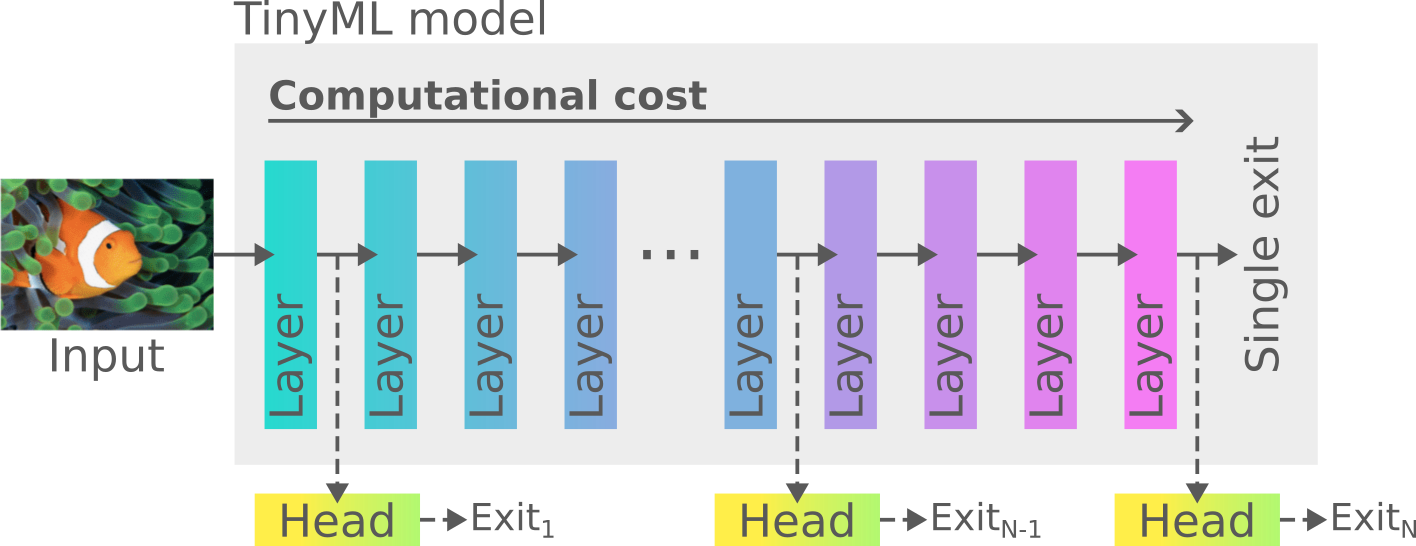}
\caption{A single-exit TinyML model with the multi-exit execution scheme.}
\vspace{-1.2em}
\label{fig:intro_image}
\end{figure}

However, TinyML models have grown in size and complexity, reaching tens of billions of MAC operations~\cite{leeVisionTransformerModels2024, MaxViT}, demanding more lightweight and energy-efficient execution schemes that move beyond the traditional approach of \textit{one fixed, optimized model always executing the same number of operations regardless of input complexity}.
Adaptive execution schemes, such as early-exit, are well-established techniques for selectively reducing the computational burden of AI workloads~\cite{teerapittayanon2016branchynet}.
By introducing intermediate classification heads, called exits, in deep neural networks (DNNs), as shown in Figure~\ref{fig:intro_image}, ``easy'' inputs can halt their processing early, while ``hard'' ones can leverage deeper layers of the network.
Early exit techniques have been widely explored for high-end devices, such as GPUs and CPUs, in computer vision~\cite{teerapittayanon2016branchynet, Wang_2018_ECCV}, natural language processing~\cite{xin-etal-2021-berxit}, and, more recently, large language model reasoning~\cite{yang2025dynamicearlyexitreasoning}.

However, despite their potential, \textbf{early-exit approaches for resource-constrained MCUs remain marginally explored} due to the multiple orders of magnitude more stringent memory, compute, and energy constraints compared to GPUs.
In fact, adding classification heads to an AI model introduces parameter and memory overhead, which is negligible on GPU-class devices but critical on ULP MCUs with only a few 100s \SI{}{\kilo\byte} of on-chip memory.
Therefore, we contribute with \textbf{a novel exploration (from design to deployment and assessment) of the early-exit computational paradigm~\cite{teerapittayanon2016branchynet, yang2025dynamicearlyexitreasoning} applied to tiny convolutional neural networks (CNNs) running on an ULP vision-based edge MCU}.

Enabling conditional execution of layers based on intermediate features, thereby making computation proportional to a sample's difficulty, is a highly attractive property for embedded intelligent systems seeking to improve efficiency.
To this end, we first introduce architectural modifications to an initial single-exit model to extract feature tensors that achieve good accuracy at each exit for part of the input data, while keeping parameter and operation counts within bounds.
Then, we identify where and how many intermediate exits we need to introduce in the initial model.
Once the architecture is defined, we handle the training process; training a multi-exit CNN is more challenging than single-exit ones, as it requires optimizing multiple loss functions, i.e., one for each exit, preventing early exits from degrading the performance of deeper ones~\cite{scardapane2020should, kaya2019shallow}, as each loss might push the optimization process towards different minima.

We address the ImageNet-100 image classification task using the MobileNetV2 CNN~\cite{sandler2019mobilenetv2invertedresidualslinear}, which we extend with our multi-exit scheme.
Each exit comes with an overhead of $\sim$\SI{25}{\kilo\nothing} parameters, accounting for a convolutional layer, a multi-layer perceptron (MLP), and a gating mechanism based on a confidence score.
For each exit, a confidence score is compared against a tunable threshold to determine whether inference should continue, or not, to the next exit.
We analyze two training strategies for multi-exit CNNs: parallel training, which jointly minimizes the loss across all exits, and sequential training, which optimizes one exit at a time while freezing the rest of the network.
Our design results in a four-exit CNN, which we quantize and deploy on an ULP GreenWaves Technology (GWT) GAP9 multicore System-on-Chip (SoC).
Finally, we characterize accuracy, operations, power, and energy on the GAP9 using both \texttt{float16} and \texttt{int8} data types, and provide an in-depth analysis of the energy-accuracy trade-offs and a state-of-the-art (SoA) comparison.

Our results show that joint parallel training across all exits yields higher accuracy than sequential training, with a loss of only $\sim$1\% accuracy vs. the single-exit MobileNetV2, which achieves 80.5\% Top-1 accuracy.
Compared with the standard single-exit MobileNetV2 on a GAP9, our multi-exit strategy reduces the average number of operations from 313 to \SI{185}{\mega MAC}, i.e., a 41\% reduction.
When deployed on the GAP9, our multi-exit CNN has an average inference time of \SI{35.0}{\milli\second}, 29\% less than the baseline single-exit MobileNetV2, and average energy consumption per frame of \SI{1.6}{\milli\joule}, 24\% less than the baseline, while maintaining a memory footprint of only \SI{2.6}{\mega\byte}.
Compared with SoA multi-exit systems~\cite{9772720}, which we also deploy on GAP9, our approach improves computational efficiency by more than 2$\times$ by fully leveraging the SoC parallelism.
Ultimately, our work demonstrates that early-exit inference, a technique widely used on GPUs, can also be beneficial for sub-\SI{100}{\milli\watt} MCUs, advancing the SoA in adaptive TinyML and bringing us closer to long-lasting, battery-powered ULP embedded intelligent systems.
\section{Related work}

\renewcommand{\arraystretch}{0.90}

\begin{table*}[t!]
\centering
\caption{State of the art review on adaptive multi-exit methods (N.A. means \textit{not available}).}
\label{tab:related}
\resizebox{\linewidth}{!}{
\begin{tabular}{l l l l l c c c}
\toprule
\textbf{Paper} & \textbf{Early-exit criterion} & \textbf{Task} & \textbf{Dataset} & \textbf{Device} & \textbf{Operations [\SI{}{\mega MAC}]} & \textbf{Parameters  [\SI{}{\mega\nothing}]} & \textbf{AVG power [\SI{}{\watt}]}\\
\midrule

BERxiT~\cite{xin-etal-2021-berxit} & Learning-to-exit & Regression & STS-B & NVIDIA P100 & $>$10000 & $>$12 & $>$250 \\
DyCE~\cite{Wang_2025} & Per exit threshold & Classification & ImageNet-1000 & N.A. (GPU) & $>$490 & $>$3.7 & $>$100\\
Dynamic Early Exit~\cite{yang2025dynamicearlyexitreasoning} & Static threshold & Reasoning & LiveCodeBench & N.A. (GPU) & N.A. & 7000 & $>$100 \\
E2CM~\cite{9891952} & Class means & Classification & ImageNet-1000 & N.A. (GPU) & 220 & 5.3 + saved (2 MB) & $>$100 \\ 
Multi-Scale~\cite{huang2018multiscale} & Static threshold & Classification & ImageNet-1000 & N.A. (GPU) & 300 & 16.8 & $>$100\\
Predictive Exit~\cite{10.1609/aaai.v37i7.26042} & Predictive engine & Classification & CIFAR-100 & NVIDIA Jetson TX2 & $>$10000 & $>$20.6 & 7.5-15.0\\
\addlinespace

\multirow{ 2}{*}{QUTE~\cite{ghanathe2024qute}} & \multirow{ 2}{*}{Distillation + ensemble} & \multirow{ 2}{*}{Classification} & \multirow{ 2}{*}{CIFAR-10} & ARM Cortex-M4 & 24 & 0.2 & 0.03\\ 
& & & & ARM Cortex-M7 & 13 & 0.3 & 0.29\\
\addlinespace
Harvester~\cite{9772720} & Static threshold & Classification & CIFAR-10 & ARM Cortex-M0 & 0.4 & 0.1 & $\sim$0.12\\
\textbf{Ours} & \textbf{Static threshold} & \textbf{Classification} & \textbf{ImageNet-100} & \textbf{GWT GAP9} & \textbf{319} & \textbf{2.6} & \textbf{$\sim$0.09}\\
\bottomrule
\end{tabular}
}
\end{table*}

Traditional approaches to enable multi-exit execution—also referred to as \textit{early exit}—in CNNs adapt the computational cost to the complexity of each input sample by inserting intermediate output branches. 
These branches allow inference to terminate early once a predefined confidence or accuracy threshold is met. 
This section reviews prior work on multi-exit neural networks, ranging from methods targeting power-unconstrained platforms such as GPUs to approaches designed for embedded systems. 
Table~\ref{tab:related} summarizes these works, listing the deployment platform when available. 
N.A. denotes missing information, while N.A.~(GPU) indicates that the platform is GPU-based but not further specified.

\textbf{Multi-exit CNNs on GPU-class devices.}
Early-exit strategies have been explored for both regression and classification tasks.
For regression tasks, early termination is typically determined based on output stability or predictive uncertainty~\cite{xin-etal-2021-berxit}, whereas in classification tasks, exit decisions are commonly triggered using softmax confidence or entropy-based criteria~\cite{teerapittayanon2016branchynet,yang2025dynamicearlyexitreasoning}.
Several works on multi-exit classification networks propose exit-specific thresholds, either precomputed and stored in an exit controller as a lookup table~\cite{Wang_2025} or derived from class-mean distances at test time~\cite{9891952}, thereby enabling runtime trade-offs between accuracy and computational cost.
Multiple works have proposed mechanisms to estimate the exit point at the start of each inference, at which a prediction becomes reliable~\cite{10.1609/aaai.v37i7.26042}, shifting the decision process from intermediate exits to a lightweight module that runs upfront to determine how many layers need to be executed.
 
MSDNet~\cite{huang2018multiscale} computes feature maps at multiple resolutions and connects them densely across layers, enabling information to flow both in depth and across scales. 
Classifiers are attached only to the coarsest, most semantic features and reuse all coarse-scale features up to their layer, giving even shallow classifiers multi-scale information and reducing the accuracy gap to the final classifier.
However, dense connectivity significantly increases memory demands: MSDNet requires \SI{16.8}{\mega parameters}, which would severely impact performance on a constrained edge device with only a few \SI{}{\mega\byte} of memory, such as the GAP9, as tensors would need to be continuously swapped between RAM (off-chip) and the on-chip memory used for computation.
Conversely, ensemble-based uncertainty estimation improves confidence assessment at the cost of storing and executing multiple network instances per input~\cite {lakshminarayanan2017simple}, thus preventing its adoption on resource-constrained devices that cannot afford such computational overhead.
To limit the overhead of ensemble uncertainty estimation, \cite{qendro2021early} proposes aggregating predictions across multiple exits of a single network, effectively forming an ensemble over exits rather than over separate model instances.

Training multi-exit neural networks is often challenging due to gradient conflicts, where the supervision from multiple exits competes during backpropagation, leading to unstable updates and suboptimal performance across exits~\cite{9190812}.
To address this problem, Deep Feature Surgery~\cite{gong2024deep} proposes exit-specific feature partitioning and referencing.
Feature partitioning splits backbone features into shared and exit-specific components, reducing interference between exits. 
Feature referencing then allows each exit to access both its exit-specific and the shared features during the forward pass, while in backpropagation, exit-specific features are updated only by their corresponding exit, and shared features are updated by all exits.
Feature Surgery~\cite{gong2024deep} effectively reduces gradient conflicts, improving early exit accuracy by $\sim7\%$  while reducing computational cost by 50\% on average if compared to MSDNet~\cite{huang2018multiscale}. 

\textbf{Multi-exit CNNs on embedded systems.}
Early-exit mechanisms in CNNs deployed on low-power embedded devices enable dynamic reduction of both inference latency and energy consumption.
Authors in~\cite{ghanathe2024qute} develop an uncertainty quantification method on four classification tasks, i.e., CIFAR-10~\cite{cifar10}, TinyImageNet~\cite{Tiny_imagenet}, SpeechCmd~\cite{SpeechCmd}, and MNIST~\cite{6296535}.
They use two STM32 Nucleo boards, i.e., the Nucleo-32 (ARM Cortex-M4) and the Nucleo-144 (ARM Cortex-M7), with a peak power consumption of $\sim$\SI{30}{\milli\watt} and \SI{290}{\milli\watt}, respectively.
The authors deploy a shallow four-layer network for the MNIST and SpeechCmd tasks, while they employ a ResNet-8 neural network for the CIFAR-10 task.
The ResNet-8 achieves a latency per frame of 59.6-\SI{298.1}{\milli\second} on the ARM Cortex-M7 and ARM Cortex-M4, respectively, requiring 17.2 and \SI{8.9}{\milli\joule} per frame for a 32$\times$32 RGB image.
Finally, TinyImageNet uses a MobileNetV2 network, but it is not deployed because it exceeds the available memory on either platform.
While QUTE~\cite{ghanathe2024qute} appends additional lightweight classification heads at the final layer to estimate prediction uncertainty and uses early-exit blocks, only during training, to promote ensemble diversity before discarding them at inference, our approach leverages trained early exits to reduce computation at runtime.

Authors in~\cite{9772720} employ an ARM Cortex-M0 processor, powered by a \SI{5.7}{\milli\meter} $\times$ \SI{6.1}{\milli\meter} \SI{684}{\milli\joule} battery and recharged via a solar panel, to perform CIFAR-10 classification.
The system runs a two-exit CNN that requires only \SI{73.2}{\kilo\byte} of memory, with exits selected based on confidence thresholds, as in our case.
Experiments on CIFAR-10 performed in~\cite{9772720} show that the multi-exit system reduces the average number of operations per frame by 43.9\% compared to the last of the two exits, with a drop in accuracy of 3.7\% (from 69.9 to 66.2\%).
However, their system requires \SI{1.27}{\minute} to perform a single inference on the smaller computation path (i.e., first exit), which increases to \SI{4.07}{\minute} for the whole network (i.e, the second exit).
Even considering the dimension of their network ($<$\SI{0.1}{\mega\byte}) and a power consumption of $\sim$\SI{120}{\milli\watt} on the ARM Cortex-M0, their inference time leads to an energy consumption per frame of \SI{19.8}{\milli\joule} using the first exit while requiring up to \SI{63.7}{\milli\joule} for the second one.
In contrast, we design a lightweight four-exit MobileNetV2~\cite{sandler2019mobilenetv2invertedresidualslinear} tailored for constrained edge devices, in which low-complexity inputs are classified using only a small fraction of the network (the first exit requires $\sim$83\% fewer operations than the deepest exit), achieving up to 80\% accuracy on the ImageNet-100 classification task.
Furthermore, our network achieves, on average, \SI{17.2}{MAC} operations per cycle, i.e., more than 2$\times$ increase compared to~\cite{9772720} due to higher parallelization.
\section{Background}

\subsection{Baseline CNN}

\begin{figure}[t!]
\centering
\includegraphics[width=0.99\linewidth]{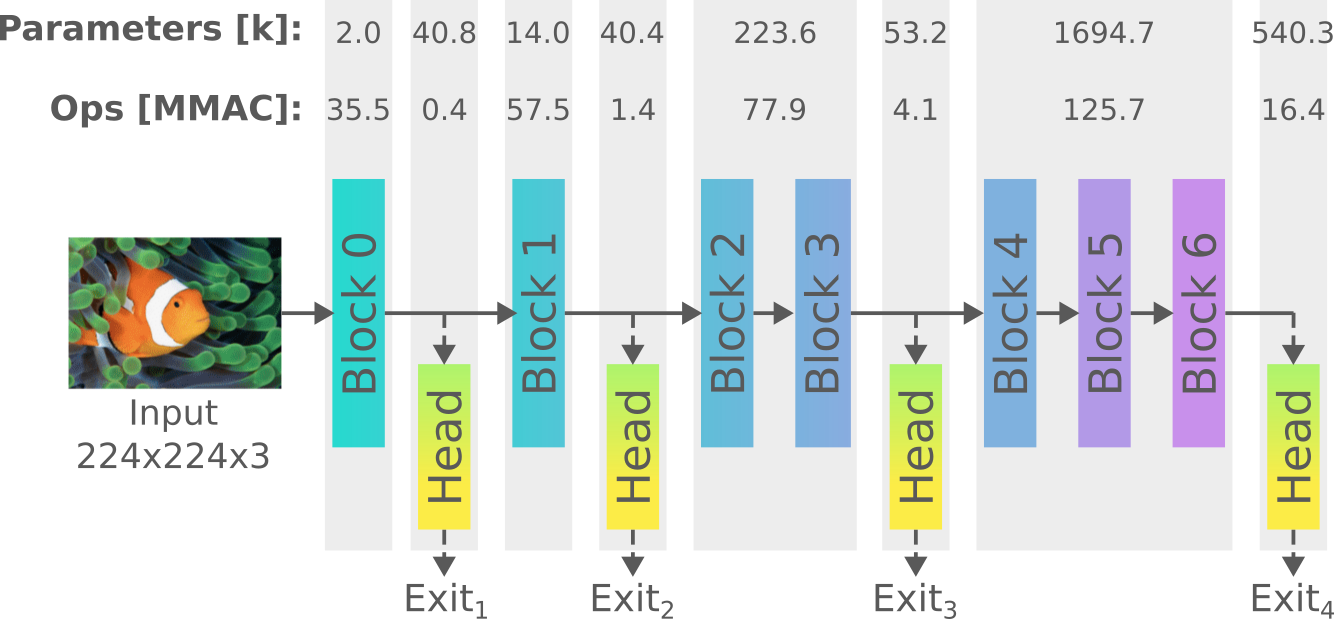}
\caption{Our MobileNetV2-based multi-exit architecture, reporting weight memory and operations per component.}
\label{fig:mobilenetv2}
\end{figure}

We employ MobileNetV2~\cite{sandler2019mobilenetv2invertedresidualslinear} as our baseline CNN, with an input size of 224$\times$224 and a width multiplier of 1.0, which scales the number of output channels at each layer.
The first layer is a 3$\times$3 convolution (from 3 to 32 channels), followed by seven inverted residual bottleneck stages with 16, 24, 32, 64, 96, 160, and 320 channels, and finally a last $1 \times 1$ convolution that projects the features to 1280 channels.
The classification head consists of batch normalization, ReLU6 activations, global average pooling, and a fully connected layer followed by a softmax; this configuration is the MobileNetV2 baseline (exit~4) reported in Figure~\ref{fig:mobilenetv2}.

We train our networks on the ImageNet-100 dataset, a widely used subset of ImageNet-1k comprising 100 classes, using a single Nvidia RTX~4080 GPU. 
The dataset contains 1300 images per class, split 80\%/20\% into training and validation, and we test on the separate ImageNet-100 test set (50 images per class).
All RGB input images are resized to $224 \times 224$ pixels and normalized to the $[0,1]$ range.
We train for 400 epochs (batch size 64) using stochastic gradient descent with momentum 0.9, weight decay $4 \cdot 10^{-5}$, and an initial learning rate of 0.01 decayed by 10× at epochs 200 and 300.

\begin{figure}[t!]
\centering
\includegraphics[width=1.0\columnwidth]{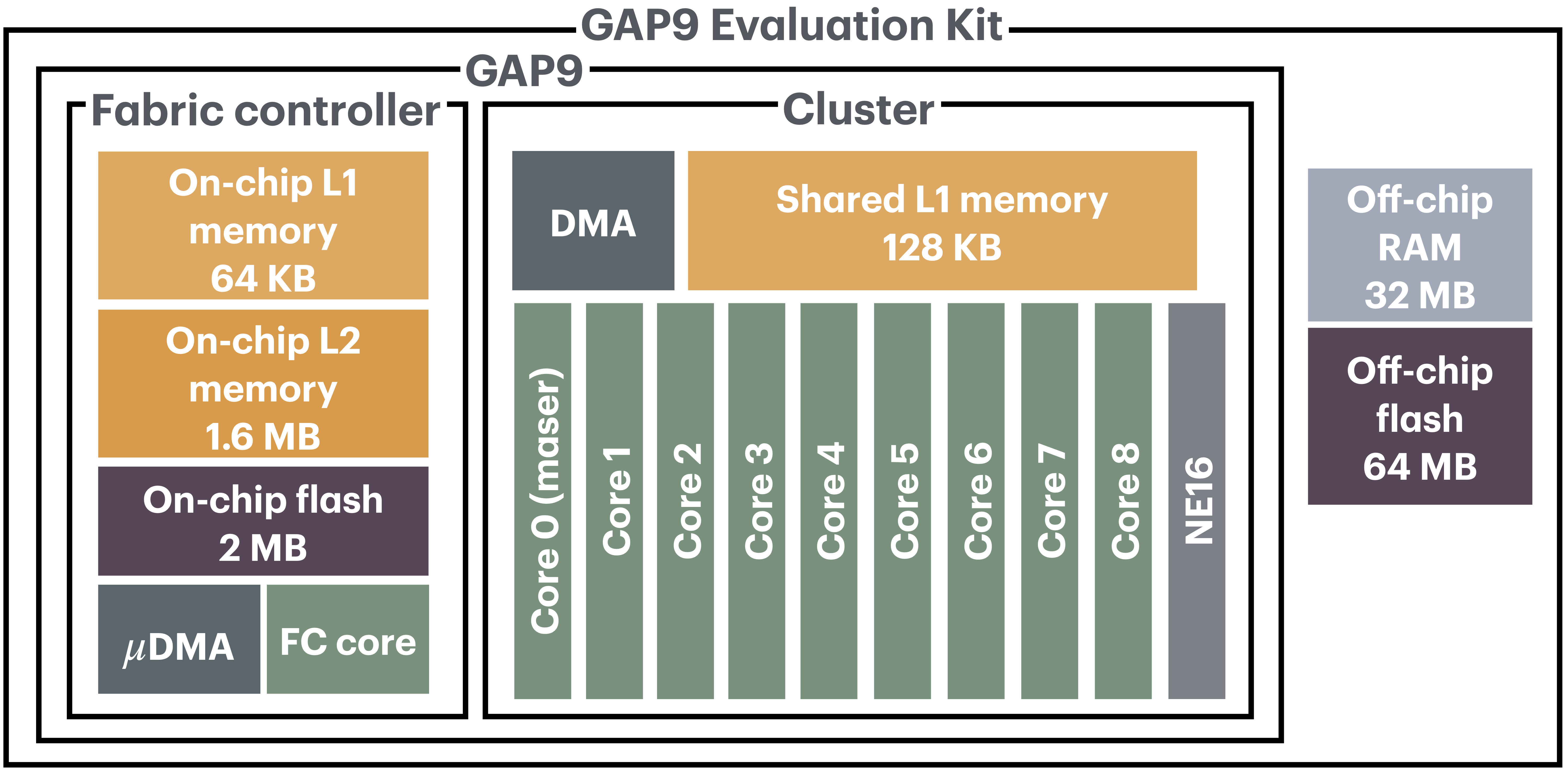}
\caption{GAP9 evaluation kit block diagram.}
\label{fig:gap9evkit}
\end{figure}

\subsection{Embedded platform}

Our target MCU, detailed in Figure~\ref{fig:gap9evkit}, is the GWT GAP9 SoC, an ULP SoC designed for parallel execution of compute-intensive workloads such as CNNs.
GAP9 features a heterogeneous architecture organized into two frequency and voltage domains: the Fabric Controller (FC), with a single RISC-V core that manages peripherals and orchestrates execution, and the Cluster (CL), including nine general-purpose RISC-V cores, four mixed-precision
floating-point units (FPUs) (FP16/BF16/FP32), and the NE16 hardware accelerator.
The nine cores of the CL deliver up to \SI{15.6}{\giga Op/\second}, while NE16, a hardware engine optimized for \texttt{int8} linear algebra operations, has a peak performance of \SI{150}{\giga Op/\second}. 
As a result, convolutional layers are executed more efficiently, with substantial gains in throughput and energy efficiency for quantized CNNs.

GAP9 integrates \SI{128}{\kilo\byte} of shared L1 scratchpad, \SI{1.6}{\mega\byte} on-chip L2 SRAM, and \SI{2}{\mega\byte} on-chip flash, while the GAP9 Evaluation Kit (Figure~\ref{fig:gap9evkit}) extends these resources with off-chip \SI{32}{\mega\byte} HyperRAM and \SI{64}{\mega\byte} flash.
Data movement across this memory hierarchy is managed by a direct memory access (DMA) engine, which offloads transfer operations from the cores. 
In addition, GAP9 integrates a $\mu$DMA subsystem that efficiently streams data between peripherals and memory without involving the cores.
\section{System implementation}

\subsection{Adaptive multi-exit CNN}

We design a multi-exit architecture based on MobileNetV2~\cite{sandler2019mobilenetv2invertedresidualslinear} for the ImageNet-100 classification task~\cite{ILSVRC15}, depicted in Figure~\ref{fig:mobilenetv2}, where the \emph{Blocks} correspond to the CNN backbone. 
Compared to the baseline CNN architecture, our network introduces three additional intermediate classification heads (four exits in total), enabling adaptive inference: simple inputs can be classified at shallower depths, while harder samples can propagate deeper in the network, exploiting additional learnable parameters that refine the output.

We introduce four exits, placing the three additional intermediate heads after bottleneck stages 2, 4, and 5 (with 24, 64, and 96 output channels, respectively), while the fourth exit coincides with the original MobileNetV2 output after stage 7 (320 channels).
The number of exits and their placement are guided by two goals: sampling the accuracy--computation trade-off as evenly as possible, and keeping the additional heads' cost negligible.
The four exits require approximately 11\%, 30\%, 55\%, and 100\% of the total MAC operations (35.9, 94.4, 175.0, and \SI{313.0}{\mega MAC}), spanning a wide range of computational budgets without clustering the operating points; together they form a well-distributed Pareto front, letting the user select the configuration that best fits a given constraint, e.g., target throughput, average accuracy, or battery lifetime.
We do not place an exit after the first bottleneck stage, whose features (16 channels at 112$\times$112 resolution) are computationally expensive to process and would still require pooling or additional convolutions before classification.
We also cap the number of exits at four, since these already cover the full compute range; additional heads would add memory and make the training process more challenging to converge.
In total, the three added heads cost only +4\% parameters and +1.9\% MACs over the single-exit baseline.
Figure~\ref{fig:mobilenetv2} reports the parameters and computational cost of each network stage and exit.
We also evaluate a single-exit setting with the same architecture, where predictions are always taken from a fixed exit $i \in \{1,2,3,4\}$ with the remaining heads disabled (no memory or compute cost), whereas the multi-exit model keeps all four heads active.

Training multiple exits instead of a single one can lead to suboptimal loss minimization and degraded performance on the classification task.
For this reason, we first train four single-exit baseline models, each corresponding to a truncated MobileNetV2 backbone terminating at exit~$i$.
We then compare these baselines against our multi-exit network trained using two different training strategies:

\textbf{Sequential multi-exit training.} 
In this approach, the exits are trained one at a time, starting from the final exit.
First, the backbone and the final classification head (exit~4) are trained together as a standard single-exit CNN (400 epochs).
Once this stage is completed, the backbone and exit~4 are frozen.
The remaining exits are then trained one at a time (exit~3, exit~2, and exit~1) for an additional 20 epochs each, while keeping the backbone and all previously trained exits fixed.
This procedure prevents gradient interference between exits and enables each classifier to be trained independently.

\textbf{Parallel multi-exit training.} 
In this paradigm, all four exits are trained jointly by minimizing the sum of their individual cross-entropy losses.
This joint optimization encourages cross-exit knowledge transfer and consistent feature learning, improving the accuracy of intermediate classifiers. 
Although this approach increases training complexity and may, in principle, introduce gradient conflicts, we observe no such effects in our experiments, even after 400 training epochs.

For all exits, we employ cross-entropy loss during training.
At test time, we employ an early-exit mechanism, described in Algorithm~\ref{alg:early_exit}, which enables the model to dynamically adapt its computational cost to the input complexity estimated by the confidence of each exit on the predicted class~\cite{teerapittayanon2016branchynet}. 
Changing the confidence threshold $\tau$ provides a trade-off between computational efficiency and accuracy, allowing precise control of the average inference cost on the target hardware.

\begin{algorithm}[t]
\caption{Confidence-based Early Exit Inference}
\label{alg:early_exit}
\begin{algorithmic}[1]
\For{$i \in \{1,2,3,4\}$}
    \State $c_i \gets \max_j p_i(j)$ \Comment{$p_i(j)$: softmax probability of class $j$ at exit $i$}
    \If{$c_i \geq \tau$}
        \State \Return prediction at exit $i$
    \EndIf
\EndFor
\State \Return prediction at final exit
\end{algorithmic}
\end{algorithm}

\subsection{Deployment}

The deployment on the GAP9 is performed using GAP\textit{flow} from GWT. 
Our trained PyTorch models are first converted to ONNX, then GAP\textit{flow} automatically designs tiling and tensor movement across the GAP9 memory hierarchy (L1, L2, and off-chip RAM), enabling efficient model deployment despite limited on-chip memory.
We deploy three network configurations: \texttt{float16} and \texttt{int8} on the RISC-V cores, plus an \texttt{int8} variant employing the NE16.  
To quantize the two \texttt{int8} versions, we use a subset of the training data as a calibration set to estimate the activation ranges of the tensors.
\section{Results}
\subsection{Image classification performance}

In this section, we evaluate the classification accuracy and computational cost of the proposed multi-exit MobileNetV2 on the ImageNet-100 dataset.
Accuracy is measured as Top-1 accuracy on the ImageNet-100 test set, comprising \SI{5}{\kilo\nothing} images (50 per class), while computational cost is quantified in terms of the number of average MAC operations per inference.
We report results for the single-exit baseline models and assess the average computational savings enabled by confidence-based early exiting under both proposed sequential and parallel multi-exit training strategies.

\begin{figure}[t]
\centering
\includegraphics[width=1.0\linewidth]{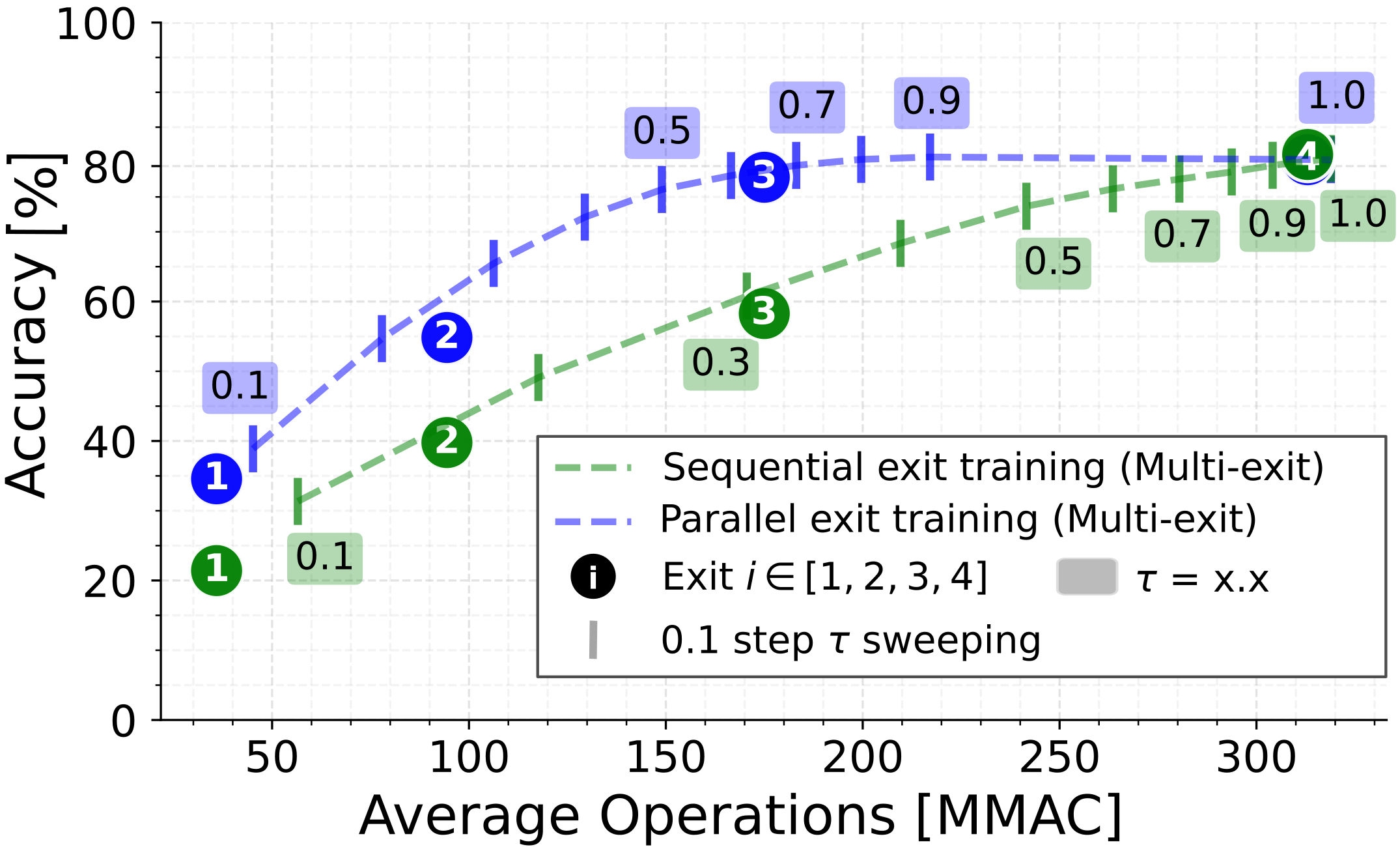}
\caption{Accuracy--operations trade-off of the multi-exit MobileNetV2 on the ImageNet-100 test set for our two training methods. The two exit 4 overlap.}
\label{fig:exit_acc}
\end{figure}

Figure~\ref{fig:exit_acc} highlights the trade-off between accuracy and computational cost, with accuracy progressively improving from the earliest to the final exit.
With sequential multi-exit training, the first three exits achieve 21.4\% (\numberincirclegreen{1}), 39.8\% (\numberincirclegreen{2}), and 58.3\% (\numberincirclegreen{3}) accuracy at computational costs of 35.9, 94.4, and \SI{175.0}{\mega MAC}, respectively, while the complete network (\numberincirclegreen{4}) reaches 80.5\% at \SI{318.9}{\mega MAC}.
With parallel training, the accuracy of the early classifiers improves substantially, reaching 34.6\% (\numberincircleblue{1}), 54.8\% (\numberincircleblue{2}), and 77.8\% (\numberincircleblue{3}) at the first three exits, while the final classifier (\numberincircleblue{4}) maintains comparable performance at 80.3\%.
Overall, both strategies achieve similar accuracy at the final exit (80.5\% sequential vs. 80.3\% parallel), whereas sequential training results in significantly lower accuracy at exits~1, 2, 3.

This difference arises from how the two strategies train the shared backbone features.
Sequential training avoids gradient conflicts because each exit is optimized independently while the backbone is partially frozen.
However, this freezing prevents shallow layers from adapting their features to exits~1--3, since the frozen representations are primarily shaped by exit~4.
In contrast, parallel training allows all exits to jointly influence the backbone via backpropagation, enabling shallow layers to learn representations useful across all exits.
\textbf{Since parallel multi-exit training yields higher accuracy at all early exits while maintaining comparable performance at the final exit, we adopt it as the preferred strategy for evaluating our multi-exit neural network.
}

We now assess the benefits of dynamic early exiting at test time, using the network trained with the parallel approach. 
To quantify this effect, in Figure~\ref{fig:exit_acc} we sweep the $\tau$ in the 0--1 range, with a growing step of 0.1, and execute inference on the entire test set for each $\tau$. 
Selecting $\tau=0$ forces the execution to stop at the first exit, whereas $\tau=1$ forces the CNN execution to stop at the last exit. 
Intermediate values distribute samples across the four exits in proportion to their confidence. 
At $\tau=0.7$, early exiting reduces the average computational cost to $\sim$\SI{185}{\mega MAC}, i.e., 41\% fewer operations per frame than the last exit, with a $\sim$1\% accuracy drop.
Accepting a 4\% accuracy drop, corresponding to $\tau=0.5$, reduces the average computational cost by 52.1\% compared to always executing the final exit.  
These findings confirm that our method achieves significant average computational savings by dynamically selecting the best model exit point while maintaining accuracy comparable to full-network execution.

\begin{figure}[t!]
\centering
\includegraphics[width=1.0\linewidth]{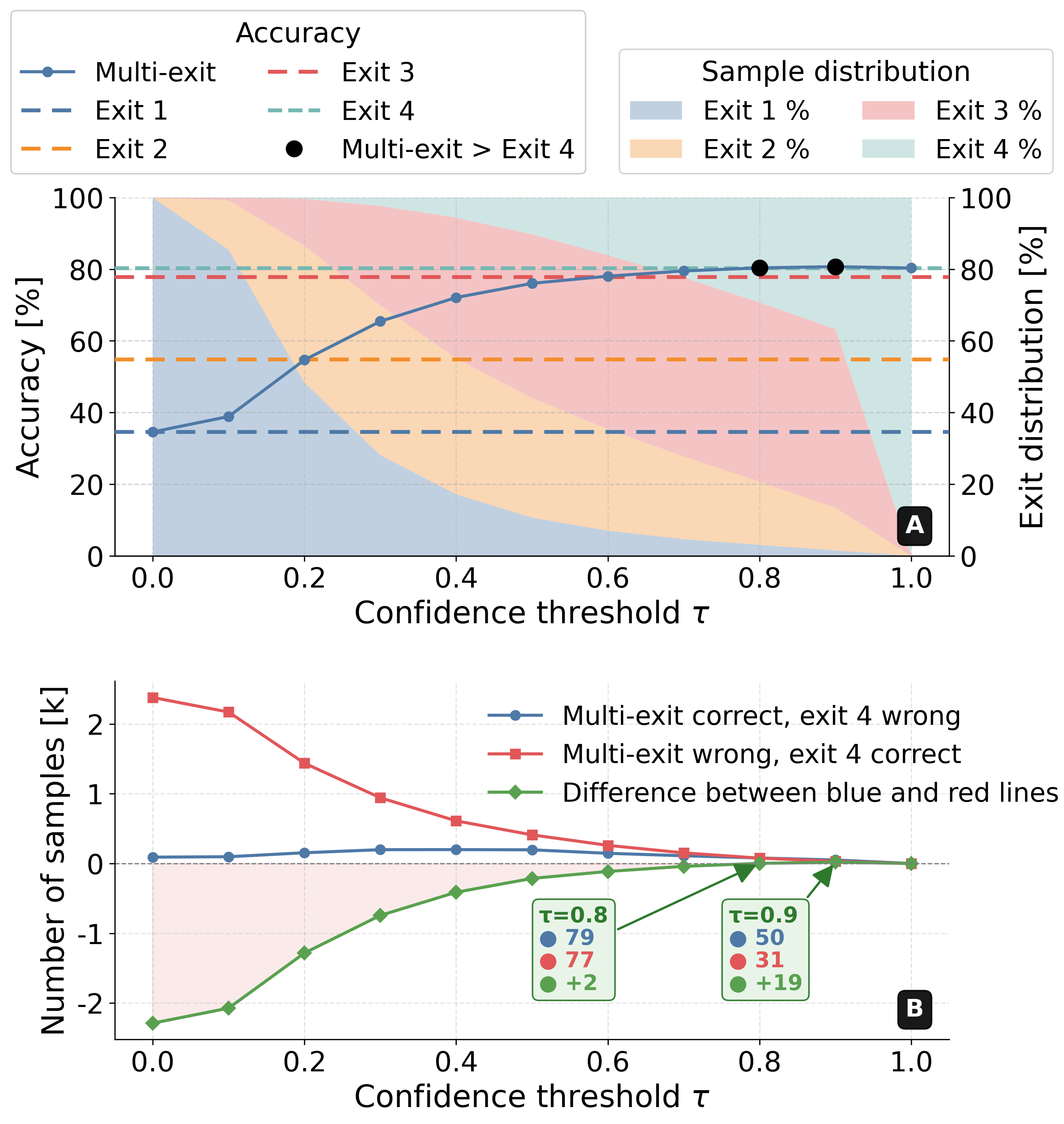}
\caption{(A) Multi-exit accuracy (left axis) and exit distribution (right axis) as a function of $\tau$. (B) Number of correct/wrong samples predicted by our multi-exit CNN vs. the single-exit model using always exit~4.}
\label{fig:corrected_degraded}
\end{figure}

Figure~\ref{fig:corrected_degraded}-A shows the exit distribution across different $\tau$ values. 
At low thresholds, most samples exit early, while increasing $\tau$ progressively routes more samples toward deeper exits.
For instance, at $\tau=0.7$, the samples are distributed across all four exits, with 22\% and 51\% of the inputs processed up to exits~3 and~4, respectively, yielding an accuracy within 1\% of always exiting at exit~4.
Notably, at $\tau=0.9$, the multi-exit model achieves 80.7\% accuracy, surpassing the single-exit baseline by 0.4\%.
Figure~\ref{fig:corrected_degraded}-B explains this counter-intuitive result. 
For every value of $\tau$, we report: the number of samples correctly classified by the multi-exit approach but misclassified by the single-exit 4 baseline (i.e., the model that always uses exit 4); and the number of samples misclassified by the multi-exit approach but correctly classified by the single-exit 4 baseline. 
At $\tau=0.9$, 50 samples are in the former group and 31 in the latter; the multi-exit model therefore yields 19 additional correct classifications w.r.t. the baseline (+0.4\% accuracy gain).
This suggests that deeper layers do not always improve predictions: some samples correctly classified at earlier exits become misclassified at the final exit, and early exiting can partially mitigate this effect.

\subsection{Embedded system performance}

\begin{table*}[t!]
    \Large
    \centering
    \caption{Deployment results of the multi-exit MobileNetV2 on GAP9 with \texttt{float16} (CL), \texttt{int8} (CL), and \texttt{int8} (NE16), reporting memory, throughput, energy, and accuracy per each exit and for our multi-exit neural network.}
    \label{tab:gap9_results}
    \resizebox{\textwidth}{!}{%
    \begin{tabular}{lcccccccccc}
    \toprule
    \multirow{2}{*}{\textbf{Architecture}} & \multicolumn{3}{c}{\textbf{Parameters memory footprint [\SI{}{\mega\byte}]}} & \multicolumn{3}{c}{\textbf{Average latency per frame [\SI{}{\milli\second}]}} & \multicolumn{3}{c}{\textbf{Average energy per frame [\SI{}{\milli\joule}]}} & \multirow{2}{*}{\textbf{Accuracy [\%]}} \\
    \cmidrule(lr){2-4} \cmidrule(lr){5-7} \cmidrule(lr){8-10}
     & \texttt{float16} & \texttt{int8} (CL) & \texttt{int8} (NE16) & \texttt{float16} & \texttt{int8} (CL) & \texttt{int8} (NE16) & \texttt{float16} & \texttt{int8} (CL) & \texttt{int8} (NE16) & \\
    \midrule
    \textbf{Exit 1} & 0.1 & $<$0.1 & $<$0.1 & 64.8 & 35.3 & 22.2 & 2.5 & 1.7 & 0.9 & 34.6 \\
    \textbf{Exit 2} & 0.1 & $<$0.1 & $<$0.1 & 128.0 & 79.1 & 26.8 & 5.1 & 3.5 & 1.2 & 54.8 \\
    \textbf{Exit 3} & 0.6 & 0.3 & 0.3 & 138.5 & 89.1 & 37.5 & 6.2 & 4.3 & 1.6 & 77.8 \\
    \textbf{Exit 4} & 5.0 & 2.5 & 2.5 & 207.9 & 120.5 & 49.3 & 9.3 & 6.0 & 2.1 & 80.3 \\
    \midrule
    \textbf{Multi-exit $\tau=0.5$} & 5.2 & 2.6 & 2.6 & 124.3 & 75.6 & 31.7 & 5.8 & 3.9 & 1.4 & 76.1 \\
    \textbf{Multi-exit $\tau=0.7$} & 5.2 & 2.6 & 2.6 & 138.9 & 85.5 & 35.0 & 6.5 & 4.4 & 1.6 & 79.5 \\
    \textbf{Multi-exit $\tau=0.9$} & 5.2 & 2.6 & 2.6 & 153.1 & 94.5 & 38.6 & 7.2 & 4.8 & 1.7 & 80.7 \\
    \bottomrule
    \end{tabular}
    }
\end{table*}

In this section, we present a detailed performance assessment of our multi-exit MobileNetV2 deployed on the GAP9 SoC. 
All experiments are performed on the GAP9 evaluation kit (Figure~\ref{fig:gap9evkit}) with the SoC operating at its maximum frequency of \SI{370}{\mega\hertz} and with a voltage of \SI{0.8}{\volt}.  
We consider three deployment variants of the network: \texttt{float16} running on the general-purpose RISC-V cluster (with FPU support), an \texttt{int8} quantized version running on the cluster, and an \texttt{int8} quantized version that exploits the NE16 accelerator. 

\paragraph{\textbf{Single-exit CNNs inference performance.}}
Table~\ref{tab:gap9_results} reports memory footprint, latency, throughput, and energy consumption required for each inference for the four single-exit CNNs under the three deployment variants.
The \texttt{float16} network is the most memory-intensive, requiring up to \SI{5.0}{\mega\byte} to process inputs through exit~4.
In this configuration, the inference latency spans from \SI{64.8}{\milli\second} (exit 1) to \SI{207.9}{\milli\second} (exit 4), corresponding to a throughput of 15.4, 7.8, 7.2, and \SI{4.8}{frame/\second} when the network always exits at exit 1, 2, 3, and 4, respectively. 
Energy consumption per inference spans from \SI{2.5}{\milli\joule/frame} to \SI{9.3}{\milli\joule/frame}.
The network quantized to \texttt{int8} and executed on the cluster reduces the peak memory footprint to \SI{2.5}{\mega\byte} (exit 4), with latencies between \SI{35.3}{\milli\second} and \SI{120.5}{\milli\second}, i.e., 28.3, 12.6, 11.2, and \SI{8.3}{frame/\second} across the four exits.
Energy consumption decreases accordingly, ranging from \SI{1.7}{\milli\joule/frame} at exit 1 up to \SI{6.0}{\milli\joule/frame} at exit 4.

Leveraging the NE16 accelerator provides the best performance-energy trade-off.
For the same \texttt{int8} model, the memory footprint remains \SI{2.5}{\mega\byte}, while latency further decreases to 22.2, 26.8, 37.5, and \SI{49.3}{\milli\second} for exits 1 to 4.
This translates to 45.0, 37.3, 26.7, and \SI{20.3}{frame/\second}, with energy consumption between \SI{0.9}{\milli\joule/frame} and \SI{2.1}{\milli\joule/frame}.
These results highlight the effectiveness of NE16 in improving both throughput and energy efficiency.
Overall, multi-exit designs provide a practical means of offering multiple operating points for energy consumption and throughput on the GAP9 SoC.

\paragraph{\textbf{Inference performance with dynamic multi-exit.}}
In Table~\ref{tab:gap9_results}, we evaluate inference performance with confidence-based early exiting and compare it to the exit~4 CNN, i.e., the most accurate and computationally expensive MobileNetV2 configuration.
The \texttt{float16} multi-exit model requires \SI{5.2}{\mega\byte}, with a \SI{0.2}{\mega\byte} overhead over exit~4, while the \texttt{int8} version requires \SI{2.6}{\mega\byte}, with a \SI{0.1}{\mega\byte} overhead.
With $\tau=0.5$, $0.7$, and $0.9$, the multi-exit model achieves average accuracies of 76.1\%, 79.5\%, and 80.7\%, respectively.
Using the NE16, latency is reduced to 31.7, 35.0, and \SI{38.6}{\milli\second}, with energy per frame decreasing to 1.4, 1.6, and \SI{1.7}{\milli\joule/frame}, respectively, compared to exit~4.
At $\tau=0.7$, early exiting reduces latency by up to 29\% and energy by 24\%, with accuracy loss below 1\%.
At $\tau=0.9$, latency and energy further decrease from 49.3 to \SI{38.6}{\milli\second} (–22\%) and from 2.1 to \SI{1.7}{\milli\joule/frame}, respectively, while slightly improving accuracy (+0.4\%, 80.7\% vs. 80.3\% at exit~4), as explained by the per-sample analysis in Figure~\ref{fig:corrected_degraded}-B.

Finally, both the \texttt{int8} and \texttt{float16} versions running on the cluster exhibit higher latency and energy consumption than the NE16 model.
Nevertheless, the advantages of early exiting are consistent across all deployment variants, matching the improvements observed with the \texttt{int8} model on the NE16 accelerator.
In the \texttt{float16} configuration, early exiting reduces the average latency from \SI{207.9}{\milli\second} (exit 4) down to \SI{138.9}{\milli\second} with $\tau=0.7$ (–33\%) and the energy from \SI{9.3}{\milli\joule/frame} to \SI{6.5}{\milli\joule/frame} (–30\%).
For the \texttt{int8} configuration on the cluster, latency decreases from \SI{120.5}{\milli\second} (exit 4) to \SI{85.5}{\milli\second} (–29\%), and energy from \SI{6.0}{\milli\joule/frame} to \SI{4.4}{\milli\joule/frame} (–27\%). 
These results demonstrate that our method consistently improves computational efficiency regardless of execution unit (cluster or NE16 accelerator), or numerical precision (\texttt{float16} or \texttt{int8}).

\begin{figure}[h!]
\centering
\includegraphics[width=1.0\linewidth]{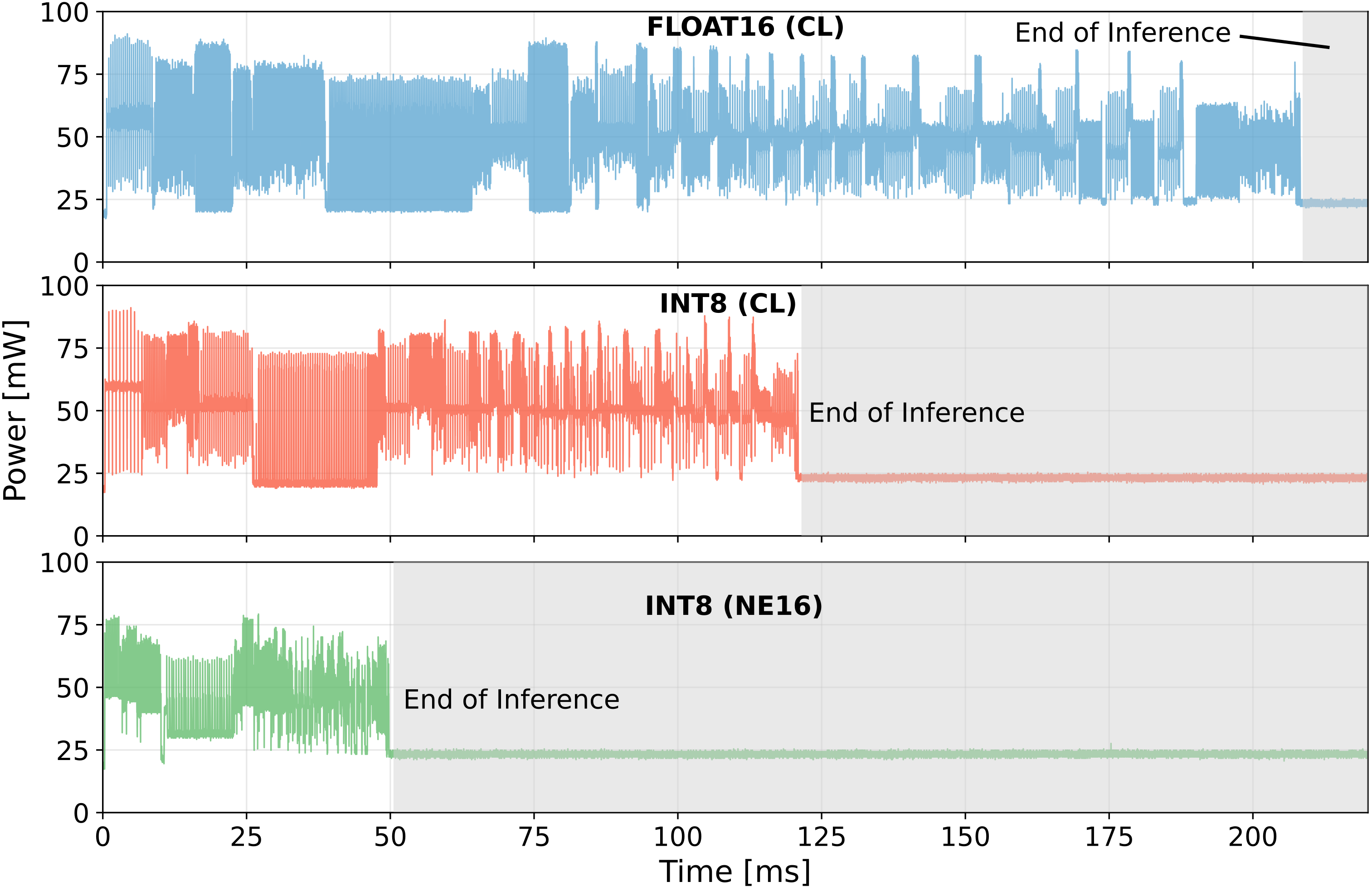}
\caption{Power waveforms of the multi-exit MobileNetV2 deployed on GAP9 when predictions are always produced at exit 4 ($\tau=1.0$).}
\label{fig:waveforms}
\end{figure}

\paragraph{\textbf{Power consumption analysis.}}
We analyze the power consumption of the proposed multi-exit CNN using a Nordic Semiconductor Power Profiler Kit II and the GAP9 evaluation kit.
Measurements account for the power drawn by both the SoC and the off-chip memories, without including the camera.
Figure~\ref{fig:waveforms} shows the measured power waveforms for the worst-case scenario, in which inference is executed up to the final exit, for the three variants: \texttt{float16}, \texttt{int8} running on the cluster, and \texttt{int8} running on the NE16 accelerator.

The \texttt{float16} implementation exhibits an average power consumption of \SI{44.83}{\milli\watt}, with a peak power of \SI{91.10}{\milli\watt}.
The \texttt{int8} cluster variant reaches a similar peak power but a slightly higher average power consumption of \SI{48.65}{\milli\watt}.
This increase is attributable to quantization, which reduces data-transfer stalls and shortens the idle periods of the cluster's processing elements.
Since idle power consumption remains constant at \SI{17.38}{\milli\watt} across all configurations, the reduced idle time results in higher average power consumption.
When leveraging the NE16 accelerator, peak power decreases to \SI{79.26}{\milli\watt}, with an average power consumption of \SI{42.14}{\milli\watt}.
These results demonstrate that the proposed multi-exit architecture operates within a \SI{100}{\milli\watt} power budget, with per-frame energy consumption ranging from \SIrange{0.9}{2.1}{\milli\joule} across best- and worst-case exit scenarios using the NE16 accelerator.

\subsection{Discussion}

This section compares our model with the four-layer CNN proposed in~\cite{9772720}, by deploying their tiny two-exit CNN on our GAP9 SoC.
When running on the NE16 accelerator, their CNN achieves inference throughput between 15 and \SI{7}{\kilo frame/\second}, depending on whether we terminate execution at the first or second exit.
However, the computational efficiency (operations per cycle) of their early exit is \SI{4.7}{MAC/cycle}, while that of their last exit is \SI{8.1}{MAC/cycle}.
In contrast, our multi-exit MobileNetV2 achieves \SI{17.2}{MAC/cycle} for full-network execution on the same platform, corresponding to a more than $2\times$ improvement in computational efficiency over~\cite{9772720}.
The low energy efficiency achieved in~\cite{9772720} stems from the minimal workload of their network, which requires at most \SI{0.4}{\mega MAC}.
Such a small computational footprint prevents effective parallel execution on both the cluster and the NE16 accelerator, as the overhead of data movement and scheduling dominates execution time.
In contrast, our deeper architecture enables higher utilization of the available hardware parallelism.

The energy efficiency of our system is further illustrated through a battery-powered deployment scenario.
We consider a millimeter-scale battery with a total energy capacity of \SI{1000}{\milli\joule}~\cite{9772720}, which acquires and processes one image every \SI{30}{\minute} over \SI{12}{\hour} of daylight.
In addition to the SoC executing the CNN, we account for a low-power camera that requires \SI{0.05}{\milli\joule} per frame and a LoRaWAN transmitter that consumes \SI{2}{\milli\joule} to transmit two bytes per classification.
Under these conditions, each frame processed by our multi-exit CNN requires an average of \SI{3.65}{\milli\joule}, with inference accounting for \SI{1.7}{\milli\joule}.
As a result, without relying on energy harvesting as in~\cite{9772720}, our system can operate for up to \SI{11.4}{\day} within the \SI{1000}{\milli\joule} battery budget.

Our evaluation targets a single architecture (MobileNetV2), dataset (ImageNet-100), and platform (GAP9). 
However, neither core component is network- or dataset-specific: the gate compares only the maximum softmax probability against a threshold, and training sums the per-exit cross-entropy losses; as such, we expect it to apply to a broad range of CNN architectures, datasets, and platforms.
\section{Conclusions}

We presented a TinyML-based adaptive image classification system that brings the multi-exit computational paradigm from GPUs to ULP MCUs.
Building upon the MobileNetV2 CNN, we designed and trained a four-exit model on the ImageNet-100 dataset, enabling dynamic early exiting via a confidence-based criterion.
We investigated two training paradigms, i.e., sequential and parallel, and found that the latter significantly improves early-exit accuracy by 13.2, 15.0, and 19.5\% at exits 1, 2, and 3, respectively, while achieving SoA accuracy at the last exit ($\sim$80\%).
Deployed on the GAP9 SoC, our solution achieves an average throughput of \SI{20.3}{frame/\second} at the deepest exit while consuming only \SI{2.1}{\milli\joule/frame}.
By tuning the confidence threshold, we show that early exiting reduces average latency (-29\%) and energy consumption (-25\%) with negligible accuracy loss ($<$1\%), and in the case of multi-exit with $\tau=0.9$ even improves accuracy (+0.4\%), while reducing the latency by 22\% compared to the single-exit MobileNetV2 baseline.
Finally, on the GAP9, our method, compared to a SoA multi-exit CNN~\cite{9772720}, improves computational efficiency by more than 2$\times$, from 8.1 to \SI{17.2}{MAC/cycle}, due to the higher level of parallelism of our system.
\bibliographystyle{IEEEtran}
\bibliography{bibliography}
\end{document}